\def\papertitle{Evaluating Neural Networks Architectures for Spring Reverb Modelling}
\def\paperauthorA{Francesco Papaleo, Xavier Lizarraga-Seijas and Frederic Font}
\documentclass[twoside,a4paper]{article}
\usepackage{etoolbox}
\usepackage{dafx_24}
\usepackage{amsmath,amssymb,amsfonts,amsthm}
\usepackage{euscript}
\usepackage[T1]{fontenc}
\usepackage[utf8]{inputenc}
\usepackage{ifpdf}
\usepackage[english]{babel}
\usepackage{caption}
\usepackage{subfig} % or can use subcaption package
\usepackage{color}
\usepackage{float}
\usepackage{booktabs}
\usepackage{comment}

\input glyphtounicode
\ninept

\newcounter{numauth}
\newcounter{listcnt}
\newcommand\authcnt[1]{\ifdefined#1 \stepcounter{numauth} \fi}

\newcommand\addauth[1]{
\ifdefined#1 
\stepcounter{listcnt}
\ifnum \value{listcnt}<\value{numauth}
\appto\authorslist{, #1}
\else
\appto\authorslist{~and~#1}
\fi
\fi}
\authcnt{\paperauthorB}
\authcnt{\paperauthorC}
\authcnt{\paperauthorD}
\authcnt{\paperauthorE}
\authcnt{\paperauthorF}
\authcnt{\paperauthorG}
\authcnt{\paperauthorH}
\authcnt{\paperauthorI}
\authcnt{\paperauthorJ}
\def\authorslist{\paperauthorA}
\addauth{\paperauthorB}
\addauth{\paperauthorC}
\addauth{\paperauthorD}
\addauth{\paperauthorE}
\addauth{\paperauthorF}
\addauth{\paperauthorG}
\addauth{\paperauthorH}
\addauth{\paperauthorI}
\addauth{\paperauthorJ}
\usepackage{times}
\newif\ifpdf
\ifx\pdfoutput\relax
\else
   \ifcase\pdfoutput
      \pdffalse
   \else
      \pdftrue
\fi

\ifpdf % compiling with pdflatex
  \usepackage[pdftex,
    pdftitle={\papertitle},
    pdfauthor={\authorslist},
    pdfsubject={Proceedings of the 27th International Conference on Digital Audio Effects (DAFx24)},
    colorlinks=false, % links are activated as color boxes instead of color text
    bookmarksnumbered, % use section numbers with bookmarks
    pdfstartview=XYZ % start with zoom=100% instead of full screen; especially useful if working with a big screen :-)
  ]{hyperref}
  \usepackage[pdftex]{graphicx}
\else % compiling with latex
  \usepackage[dvips]{epsfig,graphicx}
  \usepackage[dvips,
    pdftitle={\papertitle},
    pdfauthor={\authorslist},
    pdfsubject={Proceedings of the 27th International Conference on Digital Audio Effects (DAFx24)},
    colorlinks=false, % no color links
    bookmarksnumbered, % use section numbers with bookmarks
    pdfstartview=XYZ % start with zoom=100% instead of full screen
  ]{hyperref}
\fi
\usepackage[hypcap=true]{caption}
\title{\papertitle}

\affiliation{
\paperauthorA\,\thanks{\vspace{-3mm}}}
{\href{https://www.upf.edu/web/mtg}{Music Technology Group} \\ Universitat Pompeu Fabra\\ Barcelona, Spain\\
{\tt \href{mailto:francesco.papaleo@gmail.com}{francesco.papaleo@gmail.com} | \href{mailto:xavier.lizarraga@upf.edu}{xavier.lizarraga@upf.edu} |  \href{mailto:frederic.font@upf.edu}{frederic.font@upf.edu} }
}

\begin{document}
% more pdf-tex settings:
\ifpdf % used graphic file format for pdflatex
  \DeclareGraphicsExtensions{.png,.jpg,.pdf}
\else  % used graphic file format for latex
  \DeclareGraphicsExtensions{.eps}
\fi

%\makeatletter
%\pdfbookmark[0]{\@pdftitle}{title}
%\makeatother

\maketitle

\begin{abstract}
Reverberation is a key element in spatial audio perception, historically achieved with the use of analogue devices, such as plate and spring reverb, and in the last decades with digital signal processing techniques that have allowed different approaches for Virtual Analogue Modelling (VAM). The electromechanical functioning of the spring reverb makes it a nonlinear system that is difficult to fully emulate in the digital domain with white-box modelling techniques. In this study, we compare five different neural network architectures, including convolutional and recurrent models, to assess their effectiveness in replicating the characteristics of this audio effect. The evaluation is conducted on two datasets at sampling rates of 16 kHz and 48 kHz.
This paper specifically focuses on neural audio architectures that offer parametric control, aiming to advance the boundaries of current black-box modelling techniques in the domain of spring reverberation. 

%The results indicate that WaveNet achieves the lowest Error-to-Signal Ratio (ESR) at 16 kHz, demonstrating exceptional capability in capturing the complexities of spring reverb at this sampling rate. However, its real-time processing feasibility is limited due to a higher Real-Time Factor (RTF).
%On the other hand, the Gated Convolutional Network (GCN) presents a promising alternative, exhibiting competitive performance at both sampling rates. It maintains a significantly lower RTF, making it suitable for real-time applications with high-fidelity audio requirements.

\end{abstract}

\section{INTRODUCTION}
\label{sec:intro}
In addition to pitch, volume, timbre and tempo, spatial perception is a fundamental dimension of sound for humans. When a source emits sound, it radiates through the medium (usually air) and is received directly by the listener; this is known as direct sound. However, sound waves also reflect off surfaces in the environment before reaching the listener; these reflections arrive with a delay, attenuated in intensity and filtered in frequency, resulting in our perception of the acoustic space, which we call reverberation \cite{valimaki2012fifty}.

Throughout the history of music, the acoustics of physical locations have been used as a means to intentionally transform musical performances and provoke particular emotional states in listeners. By the early 20th century, the possibility of recording sound on a fixed medium added a new dimension unrelated to architectural acoustics that opened up new creative practices with audio effects \cite{wilmering2020history, meinema1961new}.

In terms of sound capturing, close miking reduces background noise but do not capture the acoustics of the space, resulting in very unnatural recorded material. In 1926, RCA patented reverberation chambers to overcome this limitation: a loudspeaker was placed in a room and the sound emitted was captured by a microphone placed at a distance. Later, in an attempt to achieve greater flexibility, engineers began to combine analogue electronics with mechanical systems \cite{blesser2001interdisciplinary, axon1955artificial}. This led Laurens Hammond, in 1939, to file a patent aimed at "providing an improved electric musical instrument having means of introducing a selected degree of reverberation effect into music, regardless of the acoustic properties of the place where the instrument is played" \cite{laurens1941electrical}. This technology is what we currently know as spring reverb.

After 1950 there has been a great development of analogue audio equipment, including artificial reverberation units: mainly spring, plates and tapes. These devices, which have shaped the techniques and aesthetics of contemporary music production and recording, exhibit non-linear behaviour and produce a certain degree of distortion \cite{zolzer2011dafx, valimaki2016more, gillespie2022model, bode1984history, bilbao2010perceptual}. During the 1960s, Hammond and Fender commercialized the first spring reverberation units for guitar amplifiers. Later on, professional audio equipment manufacturers such as Fisher, Fairchild and Grampian released portable units.

Since 1980, with the gradual transition to digital systems, VAM has appeared as a field of research attracting increasing attention \cite{smith1996physical, desanctis2010virtual}. The aim of VAM is to emulate the sound characteristics and behaviour of analogue audio equipment using digital signal processing methods \cite{yeh2008numerical, parker2011efficient, chowdhury2020comparison, anthon2023pywdf}. Over the years three approaches have been developed: white-box, gray-box and black-box modelling, each offer different paths to this challenge, balancing accuracy, computational efficiency, and the need for detailed circuit knowledge. With the most recent advancements in data-driven approaches, neural audio effects have emerged as a field of investigation that can lead to effective results in modelling complex effects.      

Among white-box methods, spring reverberation modelling has been approached as: parallel wave-guide structures \cite{abel2006spring}, numerical simulation techniques as finite difference schemes \cite{parker2009spring, bilbao2010virtual, bilbao2013numerical} or non-physical modelling techniques \cite{valimaki2010parametric, gardner1998reverberation}. While these approaches have shown consistent results they demand large computational resources or may struggle to model the entire set of characteristics and non-linearities of a spring reverb, to address this barrier a "DSP-informed", gray-box, has been explored in \cite{martinezramirez2020modeling}.

Whereas white-box and grey-box techniques have already been addressed for spring reverb modelling, a comparison of different black-box methods hasn't been done yet.  In previous work \cite{papaleo2023neural}, we started investigating neural network-based modelling strategies for emulating spring reverb effects. This preliminary study included a comprehensive review of relevant architectures reported in the literature, as well as an analysis of their hyperparameter configurations, loss functions and optimisation algorithms. In this study, we extended our previous work with an analysis of two publicly available datasets at different sampling rates to systematically compare five different neural network models, including convolutional and recurrent structures. We evaluate their effectiveness in capturing the unique acoustic properties of spring reverberation with quantitative metrics.

\section{BACKGROUND}

\begin{figure}[h]
\centerline{
\includegraphics[width=28em]{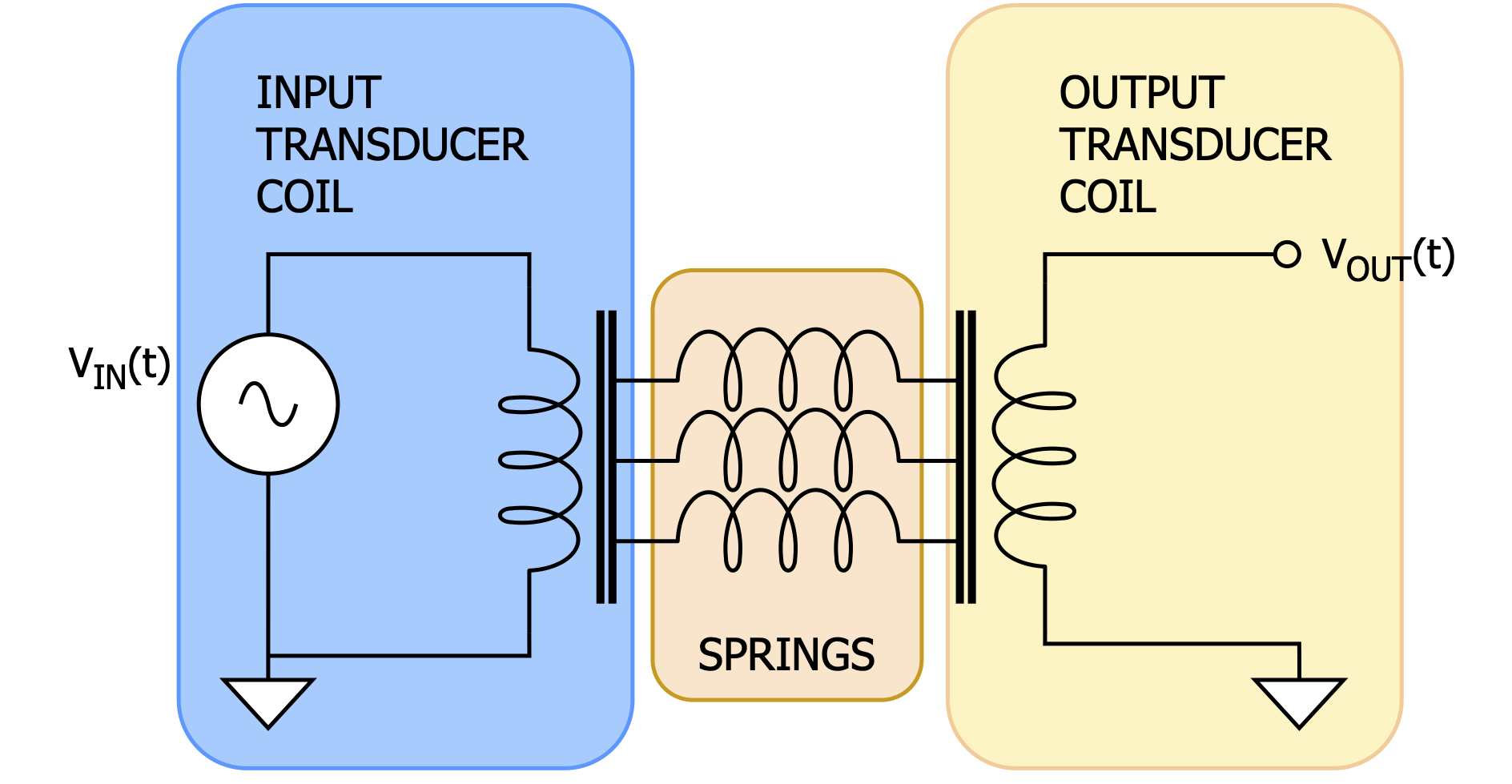}}
\caption{\it Spring tank diagram.}
\label{img:spring-tank}
\end{figure}

Figure \ref{img:spring-tank} shows the main components of a spring reverb, which consists of an input transducer, or reverb driver, that converts the incoming audio signal into mechanical vibrations by converting an electric signal in a magnetic field. These vibrations travel through one or more springs, where the propagation modes (longitudinal, transverse and torsional) cause the waves to appear delayed at the other end. These modified vibrations are then converted back into an electrical audio signal by an output transducer, or pickup. The number of springs, their diameter, length, stiffness, load and voltage are all variables that influence the salient perceptual qualities of the audio effect device \cite{abel2006spring}.

\begin{figure}[h]
\centerline{
\includegraphics[width=20em]{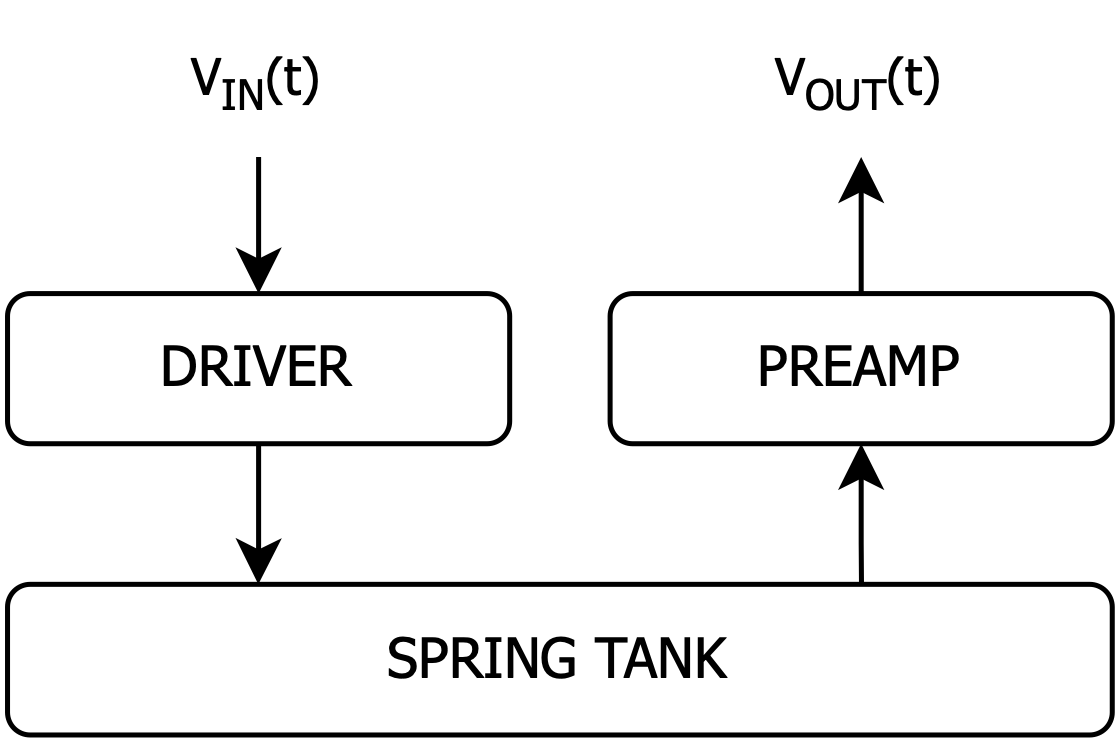}}
\caption{\it Spring reverberation unit diagram.}
\label{img:spring-reverb}
\end{figure}

The electronics of a spring reverberation unit, depicted in Figure \ref{img:spring-reverb}, is composed by a spring tank, a driver amplifier and a pre-amplifier. It is a classic compact and portable audio system that can be found in modern guitar amplifiers. Although the design of spring tanks varies depending on the manufacturer, they might be characterised by\footnote{\url{https://www.amplifiedparts.com/tech-articles/spring-reverb-tanks-explained-and-compared}}:
\begin{itemize}
    \item \textbf{Type:} it refers to the tank's length and the number of transmission springs. Length is defined in inches, where short corresponds to 9.25", miniature to 6.5" and long to 16.25". Commonly, tanks bring 2 or 3 transmission springs.
    \item \textbf{IO impedance:} input and output impedance may vary according to the manufacturer and the type. 8–200 $\Omega$ and 0.5-2k $\Omega$ are frequent ranges for input and output impedance.
    \item \textbf{Decay Time:} it indicates the time necessary to decrease in level by 60 dB (RT60). The decay time varies in function of the tank type, but it is normally on the order of 2 or 4 seconds.
    \item \textbf{Delay Time:} it is the time required for the arrival of the early reflections. 30-40 ms can be considered depending on the tank type.
    \item \textbf{Number of Springs:} Each type of tank my have 2 or 3 transmission springs. However, a transmission spring could be composed of a chain of strings altering the decay time and effect tone.
\end{itemize}

Usually the spring reverb module is interconnected within a line-out connection referenced to +4dBu or -10dBV. However the driver is not amplifying the input signal, so it is unary gain and no gain factor is applied. Its main role is adapting the impedance between a typical line-out connection, which has an output impedance from 100 to 600 $\Omega$, and the input impedance of the tank, which might have between 8-100 $\Omega$. For impedance adaptation the driver incorporates an input transformer that reduces the impedance that the tank is seeing. So the driver should be designed to provide an output impedance very low  (optimally accomplishing Rs = RL / 10) for avoiding the signal loss effect in the transmission which is determined by a voltage divider. For instance, if no adaptation is provided and we assume 1V as input signal and the lowest impedance values for Rs=100 $\Omega$ and RL=8 $\Omega$, the signal loss would correspond to -22.6dB, where us for Rs=100$\Omega$ and RL=100 $\Omega$ the loss is minimized to 6dB. However if the impedance adaptation is applied and RS is defined by RL/10 (in case RL is equal to 8 $\Omega$, Rs = 0.8 $\Omega$) the signal loss is optimized to 0.82 dB.

Other EQ parameters such as bass and treble controls can be introduced in the preamp, but they don't need to be considered as part of the reverberation time.
Usually the input is at line level signal and the  driver helps to control and keep it. Nevertheless, impedance adaptation is needed to avoid insertion loss, since the spring tank presents a low input impedance and line outputs are from 100 to 600 $\Omega$. Given the signal degradation introduced by the spring tank, the output results in a low level signal, usually about 1 to 5 mV, so the preamp applies a gain factor about 50 to 60 dB to recover a processed line level signal. The first spring reverb units are vacuum tube preamp based, it may find transistor or op-amp based units though.

The complex propagation of waves across springs causes different time of arrivals at each frequency band, sonically this introduces a nonlinear behaviour dependent on the source and filters the sound in the range from 100 Hz to 4kHz. The particular combination of electro-mechanical components and electronic parts of the spring reverb shapes a unique sound signature that, while not accurately representing acoustic reverberation, has influenced many musical traditions and styles, among others: rock, reggae, dub and funk.

% It colors the signal that's why Pitch is altered in Dataset table.
% It is a bandwidth limited audio FX, the frequency response in low frequencies is rich introducing resonances. However the coil filters frequency components below 250Hz approximetly, whereas the impedance of the spring tank in high frequencies increase drastically. 

\subsection{Spring Reverb Datasets}
Specific to spring reverb, there are only two publicly available datasets. The first one, provided by Martinez et al. \cite{martinezramirez2020modeling}, is derived from the IDMT-SMT-Audio-Effects dataset \cite{stein2010automatic}, which consists of individual 2-second dry notes that cover the common range of electric bass and guitars. The wet samples are processed through the Accutronics 4EB2C1B spring reverb tank, the dry input signal is not mixed with the output. In total, the dataset includes 624 pairs of dry and wet notes, of which test and validation samples correspond to 5\% each. All the recordings are downsampled to 16 kHz and normalized in amplitude, a fade-out is applied in the last 0.5 seconds of the recordings.  In this work, we refer to it as SpringSet.

In digital signal processing, the Nyquist-Shannon sampling theorem states that the highest representable frequency is half the sampling frequency. Therefore, the standard sampling rate of 44.1 kHz guarantees an artefact-free representation of frequencies up to 22.05 kHz. However, for deep learning in general, particularly in the early stages of development, down-sampling and up-sampling can be useful. While this reduces computing resources and training time, the cost is limited spectrum, potential artefacts and incompatibility with higher sampling rates during inference, especially in digital audio workstations (DAW).

This can be particularly crucial for reverb emulation, where high frequencies define resonances and spatial perception. Recognizing this, the Electric Guitar Effects Dataset (EGFxSet) \cite{pedroza2022egfxset}, released in late 2022, provides dry-wet pairs for various guitar effects, including spring reverb. It offers 8,970 unique, five-second annotated guitar tones with open-access availability and a sample rate of 48 kHz, and 24-bit depth. For this research, we focus on the "Clean" and "Spring-Reverb" subsets. The dataset utilizes a CR60C digital reverb emulating an Orange Crush Pro 60 amplifier with a 12" speaker. To capture spring reverb clips, the volume is set at 50\%, dry mix at 0\%, and wet mix at 100\%. The dataset isn't pre-split for training, validation and testing; each clean sample has a corresponding wet version for all 12 effects.
    
\begin{table}[ht]
  \caption{\itshape Audio features of the two datasets (mean values): Equivalent Sound Level (LEQ), Pitch (Yin algorithm) and High Frequency Content (HFC).}
  \small
	\centering
	\begin{tabular}{l c c c c }
    \toprule
            & \multicolumn{2}{c}{SpringSet} & \multicolumn{2}{c}{EGFxSet}\\
            \cmidrule(lr){2-3}\cmidrule(lr){4-5}
		Feature & dry & wet & dry & wet \\
        \midrule
	    LEQ (dB)         & -14.33          & -12.26          & -21.51        & -21.55        \\
        Pitch (Hz)       & 322.91          & 217.59          & 446.65        & 343.35        \\
        HFC             & 33.57           & 52.10           & 12.49         & 15.85         \\
        \bottomrule
	\end{tabular}
	\label{tab:features}
\end{table}

As Table\ref{tab:features} depicts, the peculiar characteristics of each spring reverb tank modify the input signal in a specific way that is reflected by the audio features computed on the two datasets with the Essentia library \cite{bogdanov2013essentia}. For both, a diminished pitch is observed due to the filtering introduced by the device. While a higher HFC between the dry and the wet signals relates to the high-frequencies resonances created by the device. As per the LEQ, in the SpringSet there's an attenuation of the processed audio that is compensated for the EGFxSet.

\begin{comment}

\begin{table}[H]
    \small
        \centering
        \begin{tabular}{l c c c}
            \toprule
            %\cmidrule(lr){1-4}
            Subset & LEQ (dB) & Pitch (Hz) & HFC \\
            \midrule

            SpringSet - dry & -14.33   & 322.91 & 33.57 \\
        SpringSet - wet & -12.26   & 217.59 & 52.10 \\
        EGFxSet - dry  & -21.51   & 446.65 & 12.49 \\
		EGFxSet - wet)  & -21.55   & 343.35 & 15.85 \\
            \bottomrule
        \end{tabular}
        \caption{\itshape Averaged audio features for SpringSet and EGFxSet.}
    \label{tab:features}
\end{table}
\end{comment}

% mention about the effect of Srping Reverb on pitch and HFC

% 

\section{DNN MODELS}
From a review of the literature on neural audio effects emerges that convolutional neural network (CNN) models can be very effective for similar tasks \cite{steinmetz2021steerable, wright2020realtime, comunita2023modelling} and at the same time there is evidence on the effectiveness of 'recurrent' type networks (RNN) \cite{wright2019realtime}. We therefore propose a comparison of five architectural patterns that revolve around these two principles of operation and, in some cases, attempt to optimise performance with hybrid-type systems \cite{martinezramirez2020deep}.

\subsection{Temporal Convolutional Network}
\begin{figure}[t]
\centerline{
\includegraphics[scale=0.35]{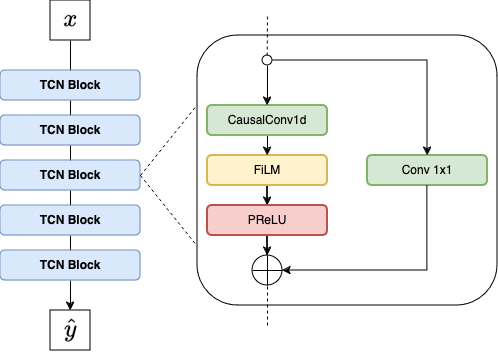}}
\caption{\it Block diagram of the TCN architecture.}
\label{img:tcn}
\end{figure}

The Temporal Convolutional Network (TCN) presented by \cite{steinmetz2021steerable} is an architecture specifically designed for processing sequential data like audio signals. As shown in Figure \ref{img:tcn}, it consists of stacked convolutional blocks with exponentially increasing dilation factors. This allows to capture both local and long-range temporal dependencies within the input without a commensurate rise in computational complexity or the number of parameters.  The first block receives the input sequence and passes it through a TCN block, transforming it into a specified number of channels. The intermediate blocks have an identical structure and simply transform the input without changing the number of channels. Finally, the last block maps the transformed features to the desired output channels. Each block employs 1D causal convolution, ensuring outputs depends only on past and current inputs.

The Feature-wise Linear Modulation (FiLM) layer introduces a mechanism for adaptive neural network behaviour through the modulation of intermediate layer outputs, conditional on external or learned information. The FiLM layer adjusts the output of convolutional layers by applying an affine transformation, whose parameters are generated dynamically based on a separate conditioning input. This allows the network to adapt its processing in a context-dependent manner, tailoring its behaviour to specific characteristics of the input signal or desired effects. Optionally, batch normalization can be integrated within the FiLM layer to introduce additional control over the feature distribution.

Following this, the architecture uses the PReLU (Parametric Rectified Linear Unit) as its activation function \cite{he2015delving}. Unlike the standard ReLU, which nullifies negative values, the PReLU allows a small gradient when the unit is inactive (i.e., for negative values). This small gradient, determined by a learned parameter, ensures that even "inactive" units can adapt during the training process, reducing the risk of neuron "death" seen in some deep network trainings with standard ReLUs. Additionally, there's a skip connection (residual) added to the output of the convolution, which helps training for deeper networks.

\subsection{WaveNet}
\begin{figure}[ht]
\centerline{\includegraphics[scale=0.35]{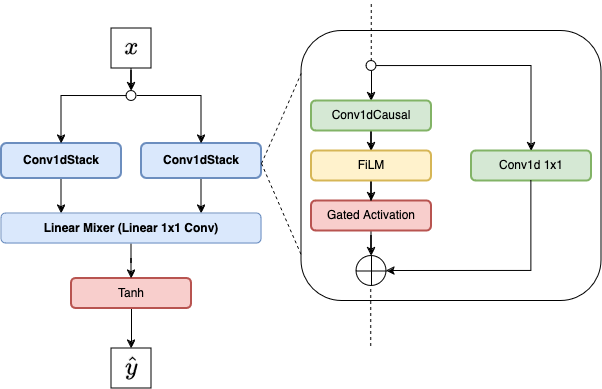}}
\caption{\label{img:wavenet}{\it Block diagram of the feed-forward WaveNet architecture.}}
\end{figure}

Figure \ref{img:wavenet} is the diagram of the second architecture implemented: a simplified, feed-forward variant of the WaveNet architecture as presented in \cite{wright2020realtime}. It utilizes stacked dilated convolutional layers to capture long-range dependencies within the audio signal. The dilation factor increases exponentially with each stack within a block. Unlike TCN's single convolutional block per layer, WaveNet employs multiple "stacks" within each block, each using a separate dilated convolution. This progressive increase in dilation allows WaveNet to capture even wider contexts without losing resolution, this comes at the cost of a higher computational complexity. 

\subsection{Gated Convolutional Network}
\begin{figure}[ht]
\centerline{\includegraphics[scale=0.35]{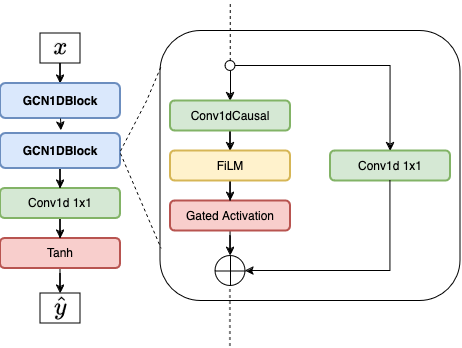}}
\caption{\label{img:gcn}{\it Block diagram of the GCN architecture.}}
\end{figure}

The third convolutional architecture is the Gated Convolutional Network (GCN), shown in Figure \ref{img:gcn} and introduced by Comunità et al. \cite{comunita2023modelling}. Similar to the other convolutional networks, this architecture can be briefly described as a TCN that leverages the power of dilated convolutions combined with gating mechanisms to capture long-range dependencies without significantly increasing computational costs. Dilation increases exponentially with the layer depth, after reaching the maximum dilation at the final layer, it resets and starts again, marking the beginning of a new block. Each block features a single causal convolutional layer (Conv1dCausal) followed by a FiLM conditioning layer and a gated activation function. The gated mechanism within the GCN modulates the output of the convolutional layers, each one has two convolutional operations. The first operation has its output channels doubled in size, and this output is split into two equal parts. One part undergoes a Tan h activation, while the other undergoes a Sigmoid activation. The element-wise product of these two activations creates the gating mechanism, effectively controlling the flow of information through the network. To maintain consistent tensor sizes throughout the network and enable residual connections, zero-padding is added at the beginning of the tensor after the gated operation. A mixing convolution of kernel size 1 follows the gating mechanism to control the number of output channels. The layer’s output is a sum of the original input (residual connection) and the result from the mixing convolution.

\subsection{Long Short-Term Memory}
Long Short-Term Memory (LSTM) networks are a type of RNNs introduced to address the problem of exploding and vanishing gradients via a gating mechanism. This mechanism allows for selective memory control, enabling the network to learn long-term dependencies in data. Figure \ref{img:rnn} shows the LSTM used is inspired by the one used in \cite{wright2019realtime} with some variations. At the input stage, the convolutional layer, enhanced with a ReLU activation and max pooling, serves as the initial feature extractor, preparing the input for the RNN layer by highlighting important temporal features and reducing dimensionality. The RNN cell offers a mechanism to manage information flow within the network, making it efficient for learning dependencies in time-series data. The FiLM layer again plays a critical role by modulating the GRU outputs according to conditional inputs, offering a mechanism for the model to adjust its behaviour based on external signals. An optional skip connections is incorporated to preserve information across layers and improve learning dynamics. 

\subsection{Gated Recurrent Units}

\begin{figure}[ht]
\centerline{\includegraphics[scale=0.35]{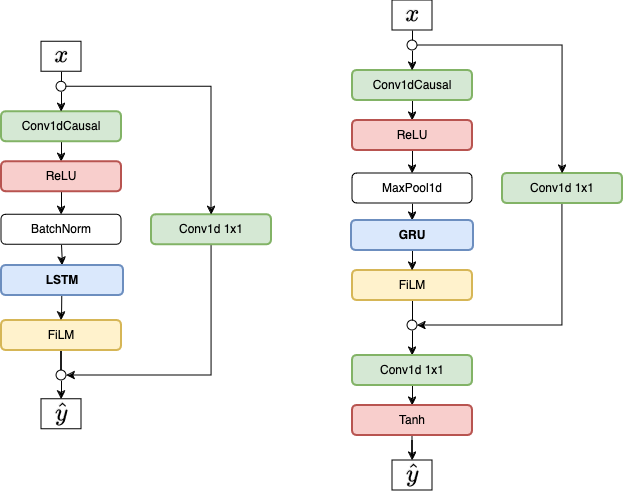}}
\caption{\label{img:rnn}{\it Block diagrams of the RNN architectures: the LSTM (on the left) and the GRU (on the right).}}
\end{figure}

Similar to LSTMs, Gated Recurrent Units (GRUs) are RNNs, they have a simpler architecture with only two gates: the update gate and the reset gate. The first one controls how much of the previous cell state is combined with the current input, and the second controls how much of the previous cell state is discarded. GRUs are generally less computationally expensive than LSTMs, but they may not be as effective at learning long-term dependencies. Figure \ref{img:rnn} shows the architecture implemented for this work, the first two stages at the input are the same as the LSTM, but they are followed by a max pooling layer before the GRU. The model concludes with a convolutional output layer and a Tanh activation function.

\section{EXPERIMENTS}

In order to ensure clarity of results and ease of replicability: for this purpose, a code base is shared to reproduce all experiments with a Command Line Interface (CLI) and the relative documentation \footnote{\url{https://github.com/francescopapaleo/neural-audio-spring-reverb}}.

\subsection{Experimental Design}
The training process is implemented using PyTorch version 2.0.1 \footnote{\url{https://github.com/pytorch/pytorch/tree/v2.0.1}}, which encompasses both the training and validation loops across all models. For both datasets, 60\% of the samples are used for training, 20\% for validation and the remaining 20\% are reserved for evaluation, with random splits.
A combination of time and frequency-domain loss functions, as detailed in \cite{steinmetz2021automatic}, \cite{steinmetz2022efficient} and in \cite{yamamoto2019probability}, is widely employed for modelling audio effects. Specifically, Mean Absolute Error (MAE) \cite{martinezramirez2018endtoend}, Error-to-Signal-Ratio (ESR) \cite{wright2021neural} and Short-Time Fourier Transform (STFT) \cite{steinmetz2020auraloss}, either alone or in combination, are commonly used as loss functions in similar tasks. The early experiments conducted in this work, aimed at identifying the most effective loss function, revealed that a combined loss of Smooth L1 \cite{girshick2015fast} and Multi-Resolution STFT (MS) was the optimal solution. The overall loss is given by: 

\begin{equation}
L =  L_{SmoothL1} + L_{STFT} 
\end{equation}

The rationale behind the use of both a time and a frequency domain loss is to ensure that the model not only captures the overall structure and envelope of the audio waveform, but also respects the harmonic complexities in the spectral domain. Loss in the time domain, represented by the smoothed MAE, is more sensitive to phase discrepancies and temporal structures, while loss in the frequency domain, represented by STFT, focuses on ensuring that the spectral characteristics of the processed audio closely match those of the original \cite{martinezramirez2020deep}.

An initial learning rate of 0.01 is set in combination with the use of the Adam optimizer and Reduce LR On Plateau as scheduler with patience of 10 epochs. The chosen scheduler reduces the learning rate by one order of magnitude, after 10 epochs in which the loss function after validation hasn't improved. For data at 16 kHz, a batch size of 64 is optimal, while with the 48 kHz this has to be reduced to 16, This difference also responds to the limited computing resources available. A single Nvidia RTX A5000 with 24 GB of RAM and CUDA version 12.0 is used for the whole experimental process.

For the evaluation of model performance, we use the ESR and Multi-Resolution STFT (MRSTFT), to measure the difference in magnitude between the predicted and the target \cite{wright2020realtime,steinmetz2020auraloss}. As shown in (\ref{esr}), the ESR is computed by dividing the absolute difference between the true value and the predicted value by the magnitude of the true value. The resulting ratio indicates how much the error deviates from the true value, relative to the true value itself.

\begin{equation}
\label{esr}
L_{ESR}=\frac{\sum_{i=0}^{N-1}\left|y_i-\hat{y_i}\right|^2}{\sum_{i=0}^{N-1}\left|y_i\right|^2}
\end{equation}

where $|y|$ is the magnitude of the true value (also called target) and $|\hat{y}|$ is the prediction given by the neural network. The denominator in the ESR normalises the loss with regard to the energy of the target signal, preventing it from being dominated by the segments with higher energy. 

Equation \ref{mrstft}, shows the MRSTFT introduced by \cite{steinmetz2020auraloss} is an extension of the STFT loss that aims to improve robustness and limits potential biases.

\begin{equation}
\label{mrstft}
L_{MRSTFT}(\hat{y}, y) = \sum_{m=1}^{M} \left( l_{SC}^m(\hat{y}, y) + \alpha l_{SM}^m(\hat{y}, y) \right)
\end{equation}

Where $M$ is the total number of resolutions and $\alpha$ is a weighting factor for the log magnitude loss. $|y|$ and $|\hat{y}|$ are, respectively, the magnitude of the ground truth and of the prediction.

Since our work aims to achieve model performance suitable for a real-time application, we include  the Real-Time Factor (RTF) among the metrics used during the evaluation: the time taken by the neural network to output a prediction is divided by the length of the input signal\cite{morise2016worlda}. An RTF less than or equal to 1 indicates that the system is employable in real time, on the other hand, a value greater than 1 indicates a delay at the output.

%\begin{equation}
%\label{rtf}
%RTF=\frac{t_p}{d}
%\end{equation}

%where $t_p$ refers to the process time and $d$ to the audio duration. An RTF less than or equal to 1 indicates that the system is employable in real time, on the other hand, a value greater than 1 indicates a delay at the output.

%\section{Results \& Analysis}

%\section{Results and discussion}
\subsection{Baseline Models}
Quantitative metrics don't always fit within predefined bounds: when there's no previous work on the same data or for a similar task, it is difficult to establish without a range of possible values. Consequently, we define and test two baseline models, built to ascertain potential metric ranges, thus aiding the subsequent evaluations of the model under comparison.
\begin{enumerate}
    \item \textbf{Naive Baseline (NB)}: This model emulates a prediction that corresponds to the wet sample and a target replaced by the dry sample. The ESR is computed between the dry and wet signals of the dataset, which is equivalent to a system that given an input returns the same input. The underlying logic of this baseline is to provide a quantitative measure of the inherent variance of the target device that we are modelling.
    \item \textbf{Dummy Regressor (DR)}: It represents the upper boundary, sometimes referred to as the "topline". It makes a random prediction that is uncorrelated with the input and the target.  Metrics are computed between a target and the random tensor (equivalent to white noise) that represents this baseline.
\end{enumerate}

\begin{figure}
    \centering
    \includegraphics[scale=0.4]
    {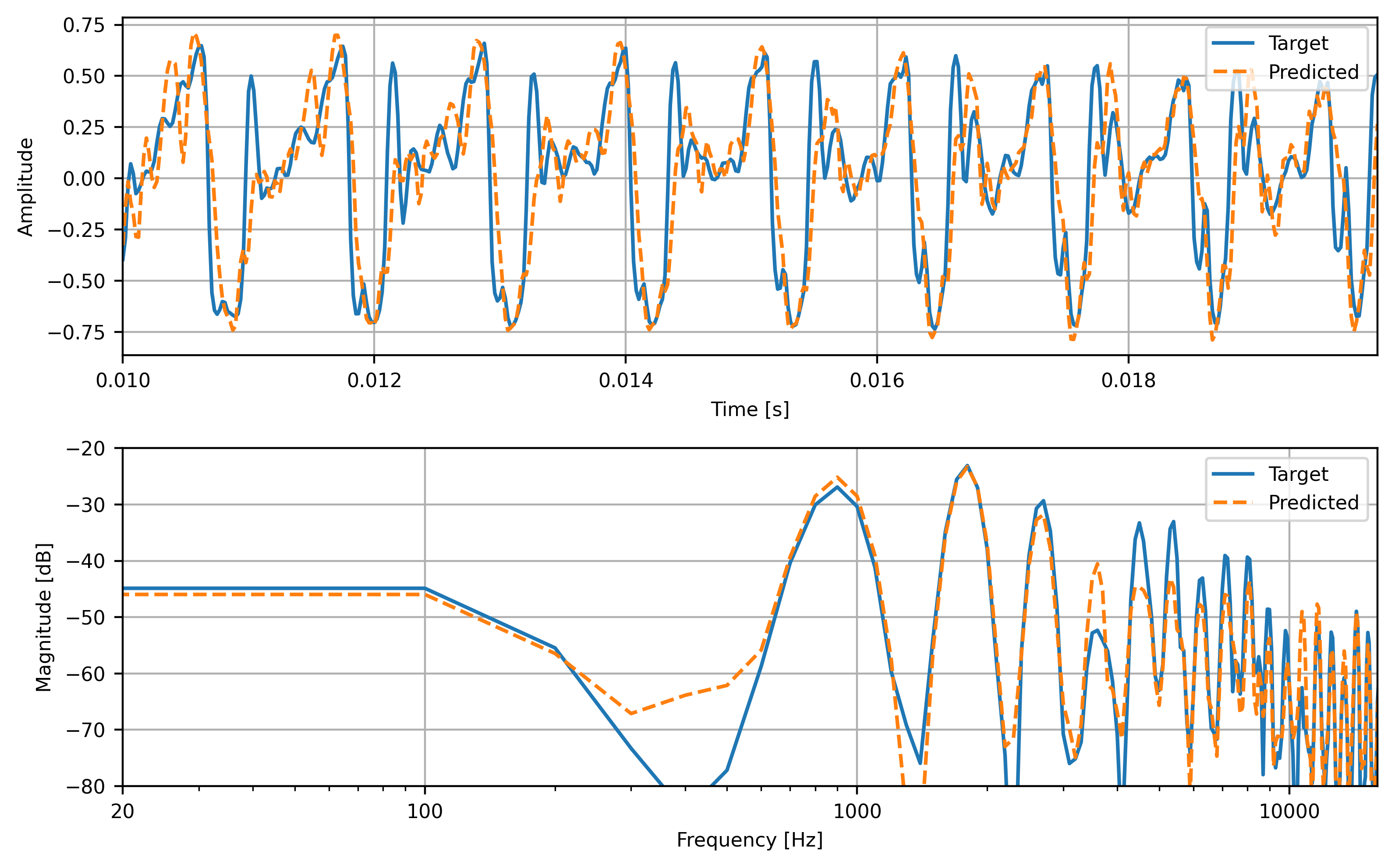}
    \caption{\it Waveforms and accumulated spectrum of the predicted and the target audio for the GCN model.}
    \label{img:waves}
\end{figure}

\subsection{Results and Discussion}
Results for both datasets are shown in Table \ref{tab:results}, while all models outperform the baseline approaches, a clear distinction emerges between the two datasets. 
\begin{table}[H]
\caption{\itshape ESR, MRSTFT (referred to as MR) and RTF on test set for all models, the lowest values highlighted in bold.}
    \small
        \centering
        \begin{tabular}{l r r r r r r}
            \toprule
            & \multicolumn{3}{c}{SpringSet} & \multicolumn{3}{c}{EGFxSet}\\
            \cmidrule(lr){2-4}\cmidrule(lr){5-7}
            Model & ESR & MR & RTF & ESR & MR & RTF \\
            \midrule

            GCN     & 0.90 & \textbf{0.96}  & 0.131 & \textbf{0.72} & \textbf{0.74} & \textbf{0.014} \\
            GRU     & 1.10 & 2.02           & 0.222   & 1.16 & 1.43 & 0.066 \\
            LSTM    & 1.06 & 2.09 & \textbf{0.114} & 1.41 & 0.99 & 0.055 \\
            TCN     & 0.80 & 1.48 & 0.144 & 1.06 & 0.93 & 0.019 \\
            WaveNet & \textbf{0.36} & 1.13 & 0.208 & 0.80 & 0.90 & 0.020 \\
            NB      & 1.40 & 1.90 & - & 1.12 & 1.84 & - \\
            DR      & 7.52 & 9.93 & - & 14.84 & 8.59 & - \\
            \bottomrule
        \end{tabular}
    \label{tab:results}
\end{table}

WaveNet achieves the lowest ESR at SpringSet, highlighting its exceptional capability in capturing the intricacies of spring reverb at 16 kHz sampling rate. Remarkably, the GCN model demonstrates competitive performance with SpringSet, and outperforms in terms of MRSTFT error also offering a potentially faster processing speed due to its simpler architecture compared to WaveNet. At the EGFxSet dataset, the GCN depicts the lowest ESR and MRSTFT, indicating its suitability for high-fidelity audio applications with a real-time processing constraint since it also exhibits the lowest RTF across both sampling rates, making it a strong candidate for real-time audio processing tasks. WaveNet while delivering exceptional performance at SpringSet has the highest RTF, indicating a potential need for optimization for real-time applications at higher sampling rates.

It is worth noticing that using the inference on CPU, it is possible to measure the time required by the models during testing to process the data and compute the RTF, although this approach does not guarantee an assessment of the deployment in a C++ framework like JUCE, it is a rough estimation of the computational demands posed by the model and allows a better tuning to improve its performance and reduce latency for a real time implementation.

\section{Conclusions}
By examining a range of convolutional and recurrent neural models, this study highlights the intricate balance between computational efficiency and the accurate reproduction of acoustic phenomena characterizing spring reverberation. Among the tested architectures, the WaveNet model demonstrated superior performance in terms of the ESR, particularly at a sampling rate of 16 kHz, underscoring the effectiveness of dilated convolutions in capturing the temporal dependencies and nuances of reverberated audio signals. Meanwhile, the GCN showed promising results at a higher 48 kHz sampling rate, and surpasses all other models in terms of MRSTFT indicating its potential for high-fidelity audio effect modelling with real-time processing capabilities.

The metrics utilized in this study provide a quantitative assessment of model performance. However, audio perception is inherently subjective, and these metrics might not always correlate with human perception. Future studies can involve more extensive perceptual evaluations, where human listeners are involved to rate the quality of the generated audio effects. Such assessments can offer invaluable insights that purely quantitative metrics might miss.

Future work, to develop more accurate neural audio effects, may involve collection of data specifically conceived for parameters learning and further experimentation with TFiLM \cite{birnbaum2019temporal}. 

\begin{comment}
\begin{figure}[ht]
\centering
\includegraphics[scale=0.4]
{gcn-99-48k.png}
\caption{\it Waveforms and spectrograms of the predicted and the target audio for the GCN model.}
\label{img:waves}
\end{figure}
\end{comment}

\section{Acknowledgments}
This research was accomplished under the project "IA y Música: Cátedra en Inteligencia Artificial y Música" (TSI-100929-2023-1), financed by Secretaría de Estado de Digitalización e Inteligencia Artificial, and Unión Europea-Next Generation EU, under the program Cátedras ENIA 2022. This paper is also based on the research conducted during a Master's Thesis on Sound and Music Computing at the Universitat Pompeu Fabra in Barcelona and the internship at Neutone Inc\footnote{https://neutone.ai/}.

%\newpage
% \nocite{*}
\bibliographystyle{IEEEbib}
\bibliography{references-clean} % requires file DAFx24_tmpl.bib

% \section{Appendix: Margin Check}
% This section shows the column margins for the text. \bigskip\newline

\end{document}